\documentstyle[14pt]{article}

\newcommand{\be}{\begin{equation}}
\newcommand{\ee}{\end{equation}}
\newcommand{\Frac}[2]{\frac{{\displaystyle  #1}}{{\displaystyle  #2}}}
\def\eop{ \framebox(6,6)\ {} }
\newcommand{\Sum}{\sum\limits}
\newcommand{\bb}[1]{\mbox{{\bf #1}} }
\newcommand{\sTr}{\mathop{\mbox{\rm sTr}}}
\newcommand{\Tr}{\mathop{\mbox{\rm Tr}}}
\newcommand{\Texp}{\mathop{\mbox{\rm Texp}}}
\newcommand{\ind}{\mathop{\mbox{\rm ind}}}
\newcommand{\sDet}{\mathop{\mbox{\rm sDet}}}
\newcommand{\Det}{\mathop{\mbox{\rm Det}}}
\newtheorem{theorema}{Theorem}
\newtheorem{lemma}{Lemma}

\title{Supersymmetric Berry Index}
\author{
{K.N.Ilinski  \thanks{E-mail: ilinski@phim.niif.spb.su} }
\\
{\small\it Institute of Spectroscopy, Russian Academy of Sciences}
\\
{\small\it Troitsk, Moscow region, 142092, Russian Federation}
\\
{G.V.Kalinin, V.V.Melezhik \thanks{E-mail: melezhik@onti.phys.lgu.spb.su}}
\\
{\small\it Institute of Physics, Saint Petersburg University,}
\\
{\small\it Universitetskiy prospect 1,
           191904 Saint Petersburg, Russian Federation}
}
\date{ }

\begin{document}

\setcounter{page}{0}
\maketitle
\vskip -9.5cm
\rightline{preprint SPbU-IP-1994-14}
\vskip 9.5cm
\thispagestyle{empty}

\begin{abstract}
        We revise the sequences of SUSY for a cyclic adiabatic
        evolution governed by the supersymmetric quantum mechanical
	Hamiltonian.
        The condition
        (supersymmetric adiabatic evolution) under which
        the supersymmetric reductions of Berry
        (nondegenerated case) or Wilczek-Zee (degenerated case)
        phases of superpartners are taking place is pointed out.
	The analogue
        of Witten index (supersymmetric Berry index) is determined.
        The final expression for new index has compact form of
$
\mathop{\mbox{\rm ind}}_{B} H =
\mathop{\mbox{\rm sDet}}U \equiv
\mathop{\mbox{\rm Det}}U^{\tau}
$,
        where $U$ is the cyclic evolution operator generated by
        supersymmetric Hamiltonian $H$ and $\tau$ is supersymmetric
        involution.

        As the examples of suggested concept of supersymmetric adiabatic
        evolution the Holomorphic quantum mechanics on complex plane
        and Meromorphic quantum mechanics on Riemann surface
        are considered.
        The supersymmetric Berry indexes for the models are calculated.
\end{abstract}

\newpage

\section{Introduction}

During last fifteen years it was proved due to the works by
E.Witten \cite{W},
A.Jaffe \cite{J}, L.Alvarez-Gaume \cite{AG} and others that the
Supersymmetric Quantum Mechanics
(SQM) and Supersymmetric Quantum Field Theory are powerful tools
to connect and
to splice geometrical and analytical substances. The Witten's
approach to
the deriving of Morse's inequalities, supersymmetric proof of
Index Theorems,
the construction of infinitedimensional analysis on the base
of Supersymmetric
Quantum Field Theory, new supersymmetric view on the complex
analysis on the
Klein surfaces \cite{BI} are some examples of such connections.
The main
technical instrument of the considerations is Witten index of
supersymmetric Hamiltonians. It is topologically stable due to
spectral properties
of supersymmetric systems and only depends on a structure of
vacuum subspace of the theory.
On the other hand Witten index can be realized as a partition function
"twisted" by the supersymmetric involution and can be
presented in the form of
a functional integral.
All these facts allow to
calculate Witten index by two ways.
The first one is to
use operator theory to investigate vacuum subspace of the Hamiltonian.
As a result geometrical nature emerges.
The second way uses Quantum Field Theory methods to compute functional
integrals.
This way leads to the analytical substances.
It is possible to refer to Chern-Gauss-Bonne Theorem and the theorem for
real-meromorphic functions on the Klein surfaces \cite{BI}
as some results of
the realization of the program. All these prompt us to look for new
topological indeces.

The main ideas of SQM approach to the calculation of new
topological indeces
can be developed in the close analogy with supersymmetric Witten index.
We have to construct quantum mechanical quantity in the framework of SQM
such that:
\begin{enumerate}
\item This quantity is calculated by means of Quantum Field Theory methods;
\item It only depends on vacuum state properties and the contributions of
superpartners with nonzero energy are vanishing.
\end{enumerate}
Some of the attempts on this direction are Supersymmetric Scattering Index
in the Supersymmetric Scattering Theory \cite{BMS} and GSQM- indeces
\cite{BIU} in
Generalized Supersymmetric Quantum Mechanics (GSQM) connected with
q- deformation of Extend supersymmetrical Quantum Mechanics \cite{IU}.
In this paper we introduce new topological supersymmetric index based on the
concept of Berry phases.

The discovery of topological phases in cyclic adiabatic
evolution \cite{B,WZ}
immediately generated a flow of papers devoted the subject. It was
very natural
to splice two topological objects: Berry phases and Witten index.
However
in the papers \cite{Ch1} it was shown that the phases of superpartners
are different in general and hence there is no possibility to invent the
topologically stable index.
To escape this difficulty we formulate the
condition which leads to the vanishing the difference of the phases and
observe this phenomenon for some well-known models.
This let us to introduce
new Supersymmetric topological index -- Supersymmetric Berry Index
(Index of Cyclic Adiabatic Supersymmetric Evolution, CASE-index).
For some
partial cases the index is an exponent of the difference of Berry phases
of zero-modes in "bosonic" and "fermionic" subspaces.
In this feature our index
is close to the Supersymmetric Scattering Index which
"calculates" the difference
of scattering phases \cite{BMS}.

The paper is organized as follows. In the section 2 we
remind the notion of topological phases for cyclic adiabatic evolution and
introduce convinient notation.
In the section 3 we prove some useful propositions for adiabatic evolution
of the supersymmetric systems. In section 4 we
formulate the conditions of Cyclic Adiabatic Supersymmetric Evolution (CASE)
and define the Supersymmetric Berry Index (CASE-index).
This index is calculated for the Holomorphic and
Meromorphic Supersymmetric Quantum
Mechanics in the section 5.
In the last section 6 we formulate conclusion
remarks on the possibility to use Supersymmetric Berry Index to derive
new index theorems.

\section{Adiabatic evolution}

Let's consider the quantum mechanical system governed by time depended
Hamiltonian $H(t),\ t\in[0,T]$. We assume that for all $t\in[0,T]$
Hamiltonian $H(t)$ has only discrete spectrum.
$E_j(t)$ denote its eigenvalues and $P_j(t)$ denote projectors on
corresponding eigenspaces. We assume that $E_j(t)$ and $P_j(t)$ are
continuous function of $t$. Denote $s = t/T$.

\begin{theorema}
\label{adiabatic}
(Adiabatic theorem) \cite{Mes}
If the adiabatic conditions
\begin{equation}
\begin{array}{l}
1^0 \quad E_{j} (s) {\neq} E_{k} (s) \quad \forall s \in [0,1],
\quad j {\neq} k ,  \\
2^0 \quad \forall j \  P_j (s) \
\mbox{is double continuously }  \\
\qquad \qquad \quad
\mbox{differentiable function of} \ s \in [0,1] ,
\end{array}
\label{Ad}
\end{equation}
take place and evolution operator  $U_T(s)$ obeys Schr\"odinger equation
$$
i\Frac{\partial}{\partial s} U_T(s) = T H (s) U_T(s)
$$
Then
\begin{equation}
\lim_{T \to \infty} U_T (s) P_j (0) =
P_j (s) \lim_{T \to \infty} U_T (s)
\end{equation}
\end{theorema}

Now we use this theorem for the deriving the
form of the adiabatic evolution
operator in terms of dynamical and geometrical phases.
Condition $2^0$ provides that $\dim P_j(t)\bb{H}$ does not depend on time $t$.
In $P_j(t)\bb{H}$ let's choose the basis
$\{\varphi_{j}^\alpha(t)\}_{\alpha=1}^{\dim P_j\bb{H}}$ and
consider wave function $\psi_{j}^\alpha(t)$ with initial condition
\begin{equation}
\psi_{j}^\alpha (0) = \varphi_{j}^\alpha (0) \ .
\end{equation}
Adiabatic theorem says that at any moment $t\in[0,T]$ the wave function
$\psi_{j}^\alpha(t)$ in the adiabatic limit is eigenfunction
of instant Hamiltonian with eigenvalue $E_j(t)$.
Therefore it can be decomposed in the basis:
\begin{equation}
\label{expand}
\psi_j^{\alpha} (t) = \sum\limits_{\alpha^\prime=1}^{\dim P_j\bb{H}}
u_{j}^{\alpha^\prime\alpha}(t)\varphi_j^{\alpha^\prime} (t)
\end{equation}
The substitution of the last expression in
Schr\"odinger equation
\begin{equation}
\label{Shred}
i\Frac{\partial}{\partial t} \psi_j^{\alpha}(t) = H (t) \psi_j^{\alpha}(t)
\end{equation}
gives
\begin{equation}
\Sum_{\alpha^\prime=1}^{\dim P_j\bb{H}}
\left(
i\dot u_{j}^{\alpha^\prime\alpha}(t) \varphi_{j}^{\alpha^\prime}(t) +
iu_{j}^{\alpha^\prime\alpha}(t) \dot\varphi_{j}^{\alpha^\prime}(t) -
E_j u_{j}^{\alpha^\prime\alpha}(t) \varphi_{j}^{\alpha^\prime}(t)
\right)
= 0
\end{equation}
The scalar product of this equation with
$\langle\varphi_{j}^\beta(t)|$
result in the equation for $u_{j}^{\beta\alpha} (t)$:
\begin{equation}
\label{u}
i\dot u_{j}^{\beta\alpha}(t) + \Sum_{\alpha^\prime}
iu_{j}^{\alpha^\prime\alpha}(t)
\langle \varphi_{j}^{\beta}(t)|\dot\varphi_{j}^{\alpha^\prime}(t)\rangle
- E_j u_{j}^{\beta\alpha}(t) = 0
\end{equation}
Let's introduce matrices
\begin{equation}
\label{notation}
U(t) = \| u_{j}^{\alpha\alpha^\prime}(t)\|  \ ,  \quad
B(t) = \| \langle
                 \varphi_{j}^{\alpha}(t)
	   |
                 \dot\varphi_{j}^{\alpha^\prime}
       \rangle \|
\ ,
\quad
E(t) = \| \delta_{j}^{\alpha\alpha^\prime}E_j(t)\| \ ,
\end{equation}
which are block-diagonal in adiabatic limit.
Blocks (numerated by $j$)
correspond to energy levels $E_j(t)$ and their dimensions are equal to
the degrees of degeneracy of the levels.
The equation (\ref{u}) can be written in the matrix form:
\begin{equation}
\dot U(t) = - (B(t)+iE(t)) U(t)
\label{ev}
\end{equation}
Taking into account the initial condition $U(0)=I$ we can write
its solution:
\begin{equation}
U(t) = \exp \left(
	          -i\int^t_0 E(s) ds
            \right)
       \Texp\left(
	          -\int^t_0 B(s) ds
            \right)
\end{equation}
In RHS of the expression the first factor has dynamical nature
the second factor is of geometrical one.
It is the second term that in the case of
cyclic adiabatic evolution gives the geometrical Berry phases.

\section{Difference of the superpartners's phases}

The main purpose of this section is to introduce supersymmetric notations,
describe the difference in Berry phases for eigenstates -- superpartners
of supersymmetric Hamiltonian $H(t)$ \cite{Ch1} and to prove some
statements we use below in section 3.

Let's suppose that the Hamiltonian $H(t)$ which manages the cyclic
adiabatic evolution is supersymmetric one i.e. the relations of the
supersymmetric quantum mechanics (SQM) take place at
any instance of $[0,T]$:
\be
\begin{array}{ll}
\tau= \tau^* = \tau^{-1} \ , \qquad   &
Q^{*}(t)=Q(t) \ ,            \\
\tau Q(t)+Q(t)\tau =0 \ ,    \qquad &
H(t)=\Bigl(Q(t)\Bigr)^2
\end{array}
\label{SQM}
\vspace{0.6cm}
\ee
Here $\tau$ is supersymmetric involution (grading operator on Hilbert
space of physical states) and $Q(t)$ is supercharge for the
supersymmetric Hamiltonian $H(t)$. In accordance with $\tau$-grading
Hilbert space splits in two subspaces ("bosonic" $\bb{H}_+$ and "fermionic"
$\bb{H}_-$ spaces) such that
\begin{equation}
\label{sum}
\bb{H} = \bb{H}_+ \oplus \bb{H}_- \ , \qquad
\tau\bb{H}_{\pm} = {\pm}\bb{H}_{\pm}
\end{equation}

On this basis the operators $\tau $, $Q(t)$, $H(t)$ can be rewritten
in the matrix form:
\begin{equation}
\array{l}
\tau =
\left(
\begin{array}{cc}
1 & 0
\\
0 & -1
\end{array}
\right) \ ,
\qquad
Q(t) =
\left(
\begin{array}{cc}
0 & q(t)
\\
q^*(t) & 0
\end{array}
\right) \ ,
\\ {} \\
H(t) =
\left(
\begin{array}{cc}
H_+(t) & 0
\\
0 & H_-(t)
\end{array}
\right) =
\left(
\begin{array}{cc}
q(t)q^*(t) & 0
\\
0 & q^*(t)q(t)
\end{array}
\right)
\ ,
\endarray
\end{equation}
with $q(t):\ D(q(t))\to \bb{H}_-$ densely defined closed operator on the
domain $D(q(t))\subset\bb{H}_+$, where $D(q(t))=D(Q(t))\cap\bb{H}_+$.
Operators $H_+(t)$ and $H_-(t)$ are called the Hamiltonian of superpartners.
With the assumption about pure discrete spectrum of the Hamiltonian $H(t)$
the relations (\ref{SQM}) provide the coincidence of the spectra of $H_+(t)$
and $H_-(t)$ at any instance exepting zero-energy levels if any.

Generally there is no connection between the zero-modes
(eigenfunctions with zero eigenvalue) of $H_+$, $H_-$ in
$\bb{H}_+$ and $\bb{H}_-$ but for nonzero-modes it is possible to do this.
However the following lemma take place.

\begin{lemma}\label{invariant}
\begin{enumerate}
\item Witten index of supersymmetric Hamiltonian $H(t)$
$$
{\ind}_W H(t) = \dim \ker \bb{H}_+(t) - \dim \ker \bb{H}_-(t)
$$
is adiabatic invariant, i.e. in the conditions of adiabatic evolution
the following equality takes place:
$$
{\ind}_W H(t) = {\ind}_W H(0) \quad \forall t \in [0,T].
$$
\item If ${\ind}_W H(0)\ne 0$ then dimensions of "bosonic" and "fermionic"
zero-mode subspaces are also adiabatic invariants, i.e.
$$
\begin{array}{l}
\dim \ker \bb{H}_+(t) = \dim \ker \bb{H}_+(0) \ ,\\
\dim \ker \bb{H}_-(t) = \dim \ker \bb{H}_-(0) \quad \forall t \in [0,T]\ .
\end{array}
$$
\end{enumerate}
\end{lemma}

{\it Proof.} Let $\dim \ker H_+(0)=m$, $\dim \ker H_-(0)=n$.
Then under the adiabatic evolution which does not change the degeneracy
of instant eigenvalues only $m$ eigenfunctions could leave zero-mode
subspace in $\bb{H}_+$ at some moment $t_0$. Due to supersymmetry arguments
this leads to the fact that $m$ eigenfunctions leave at moment $t_0$
zero-mode subspace in $\bb{H}_-$ (because as it was noted above
$\mbox{spec}H_-\setminus\{0\} = \mbox{spec}H_+\setminus\{0\}\
\forall t \in [0,T]$).
So there are two possibilities:
\begin{enumerate}
\item $m=n$ and ${\ind}_W H(0) = {\ind}_W H(t_0)=0$. It is possible to
investigate the inverse process and to consider the arriving of $2m$
nonzero-modes in zero-mode subspace. This process does not also change
Witten index. Summarizing the reflection we infer
${\ind}_W H(t) = {\ind}_W H(0) \ \mbox{for}\ \forall t \in [0,T]$.
This equality proves the Lemma \ref{invariant}
for the case of ${\ind}_W H(0)=0$.
\item $m\ne n$ and ${\ind}_W H(0)=m-n\ne 0$.
Then zero-mode subspace in the space $\bb{H}_-$ splits at the moment $t_0$
that
contradicts to the adiabacity of the evolution. So there is no possibility
for zero-modes to leave zero-mode subspace in $\bb{H}_-$ and hence in
$\bb{H}_+$.
This note results in the next equalities:
\end{enumerate}
\begin{equation}
\begin{array}{l}
\dim \ker \bb{H}_+(t) = \dim \ker \bb{H}_+(0)  \ ,\\
\dim \ker \bb{H}_-(t) = \dim \ker \bb{H}_-(0)  \ ,\\
{\ind}_W H(t)=m-n={\ind}_W H(0) \quad \forall t \in [0,T],
\end{array}
\end{equation}
which finish the proof of the Lemma \ref{invariant} \eop

Now let's consider the instant eigenfunctions
$\{\varphi_{i+}^{\alpha}\}^n_{\alpha=1}\in\bb{H}_+$,
$\{\varphi_{i-}^{\alpha}\}^n_{\alpha=1}\in\bb{H}_-$
corresponded to the instant eigenvalue
$E_i(t)\ne 0$ with the degeneracy $n$.
It is possible to choose such basis in $\bb{H}_+$, $\bb{H}_-$ that
$\{\varphi_{i+}^{\alpha}\}$, $\{\varphi_{i-}^{\alpha}\}$ are expressed
one through another:

\begin{equation}
\begin{array}{c}
q^*(t) \varphi_{i+}^{\alpha }(t) = \sqrt{E_i(t)} \varphi_{i-}^{\alpha }(t)
 \\
q(t) \varphi_{i-}^{\alpha }(t) = \sqrt{E_i(t)} \varphi_{i+}^{\alpha }(t) \ .
\end{array}
\label{conn}
\end{equation}
The eigenfunctions related as (\ref{conn}) are said to be
superpartners.

It is interesting to compare the phases gained by wave functions of
superpartners under cyclic adiabatic evolution.
It is obvious that the dynamical phases are equal.
However geometrical phases
(Berry phases) can differ. The Theorem \ref{2} shows it.

\begin{theorema} \cite{Ch1}
\label{2}
In the notations (\ref{notation})
\begin{equation}
\label{Theor2}
\begin{array}{c}
\Delta_{j}^{\alpha \alpha ^{\prime}}(t) \equiv
 B_{j+}^{\alpha \alpha ^{\prime}}(t) -
 B_{j-}^{\alpha \alpha ^{\prime}}(t) =
 \\
= \Frac{\dot E_{j}(t)}{2E_{j}(t)}\delta^{\alpha \alpha ^{\prime}} +
  \langle
         \varphi_{j+}^{\alpha }(t)
         |
  \Frac{-q{\dot q}^*}{E_{j}(t)}
         |
	 \varphi_{j+}^{\alpha ^{\prime}}(t)
  \rangle
\vspace{0.6cm}
\end{array}
\end{equation}
\end{theorema}

{\it Proof}.
The proof of the Theorem \ref{2} can be fulfilled by straightforward
calculation using the definition of the matrix B and relations (\ref{conn}).
\eop

{\bf Corollary 2.1}
Using the antihermitian property of the matrix $B_{j\pm}$:
$B_{j\pm}^{\alpha \alpha ^{\prime}}=
-\overline B_{j\pm}^{\alpha^{\prime} \alpha}$
it is possible to rewrite
$ \Delta_{j}^{\alpha \alpha ^{\prime}}(t) $
in symmetrical form
$$
\Delta_{j}^{\alpha \alpha ^{\prime}}(t)  =
\Frac{\Delta_{j}^{\alpha \alpha ^{\prime}}(t) -
\overline \Delta_{j}^{\alpha ^{\prime} \alpha }(t) }{2} =
\langle\varphi_{j+}^{\alpha }(t)|
\Frac{\dot q(t)q^*(t)-q(t){\dot q(t)}^*}{2E_j(t)}
|\varphi_{j+}^{\alpha ^{\prime}}(t)\rangle
$$
which leads to the expressions
$ \Delta_{j}^{\alpha \alpha ^{\prime}}(t) $
through the supercharge $Q(t)$:
$$
\begin{array}{r}
\Delta_{j}^{\alpha \alpha ^{\prime}}(t) =
\Frac{1}{2E_j(t)}\langle\varphi_{j+}^{\alpha }(t)|[\dot Q(t),Q(t)]
|_{\bb{H}_+}|\varphi_{j+}^{\alpha^{\prime}}(t)\rangle  =  \\
= - \Frac{1}{2E_j(t)}\langle\varphi_{j-}^{\alpha }(t)|[\dot Q(t),Q(t)]
|_{\bb{H}_-}|\varphi_{j-}^{\alpha^{\prime}}(t)\rangle
\end{array}
$$
The latter expression can be obtained by the same way starting from
eigenfunction in $\bb{H}_-$.

{\bf Corollary 2.2}
$\Tr\Delta_j$ can be expressed in supersymmetric terms
\begin{equation}
\sTr B_j(t) = \Tr \Delta_j(t) =
\Frac{1}{4} \sTr(H^{-1}(t)[\dot Q(t),Q(t)]P_j(t))\ .
\end{equation}
These relations allows us to introduce the notion of
Adiabatic Supersymmetric evolution in the next section.

\section{Cyclic Adiabatic Supersymmetric Evolution (CASE)
and its index}

{}From here on we assume that adiabatic evolution is cyclic one on $[0,T]$
i.e.
\begin{equation}
H (T) = H (0)
\end{equation}
In this case we can take instant eigenfunction which obey cyclic condition:
\begin{equation}
\varphi_i^{\alpha} (T) = \varphi_i^{\alpha} (0)
\label{cyc2}
\end{equation}

The main problem of this section is to define some class of supersymmetric
Hamiltonians which allow to construct topologically stable index on the
basis of cyclic adiabatic phase.

The usual way to calculate topological invariants into the framework
of supersymmetry is to invent the quantity in which the contributions of
superpartners with nonzero eigenvalues are cancelled and the rest depends on
vacuum subspace structure. In this way Witten index can be realized as
some trace on the space of states:
\begin{equation}
{\ind}_W H = \Tr \Bigl( \tau e^{-\beta H} \Bigr)
\quad
\forall\beta>0
\end{equation}
and its stability is a consequence of the cancelation.
We would like
to suggest the analog of the construction based on Berry phases.
On the other hand Theorem \ref{2} describes the differences for phases of
superpartners in general.
This compels us to reduce the set of
supersymmetrical Hamiltonians to extract the subset of operators for which
the difference effectivelly disappears.
At first we describe the subset
and introduce new supersymmetric index for it.
Then we show that this class
of supersymmetrical Hamiltonian  is enough wide to contain well-known
interesting examples.

{\bf Definition 1.}
\ The supersymmetrical Hamiltonian $H(t)$ admits the \\
Cyclic Adiabatic Supersymmetric Evolution (CASE) on $[0,T]$ if:
\begin{enumerate}
\item
$H(t)$ obeys the SQM algebra (\ref{SQM});
\item
$H(t)$ governs the adiabatic evolution i.e.
adiabatic condition (\ref{Ad}) take place;
\item
$H(T) = H(0)$;
\item
$\int\limits^T_0
                \sTr_{reg} (H^{-1}(t)[\dot Q(t),Q(t)])dt \equiv $
\begin{equation}
\equiv \lim\limits_{\lambda\to\infty} \int^T_0 \sTr
(H^{-1}(t)[\dot Q(t),Q(t)]E(]0,\lambda])(t))dt = 0
\label{SSev}
\end{equation}
\end{enumerate}
where $E(]0,\lambda])(t)$ is the spectral measure of the interval
$]0,\lambda]$ for the operator $H(t)$.

For the class of CASE-Hamiltonians it is possible to introduce an analog of
supersymmetric Witten index and index of Supersymmetric Scattering Theory
\cite{BMS}.
It is Supersymmetric Berry Index.

{\bf Definition 2.}
Let's $H(t)$ is CASE-Hamiltonian on $[0,T]$. Then its Supersymmetric Berry
Index is defined by the following relation:
\begin{equation}
{\ind}_B H = {\sDet}_{reg} U(T) \equiv
\lim_{\lambda \to \infty} \Det U^\tau (T)
\left. \over \right|_{E([0,\lambda])(t)\bb{H}}
\end{equation}
{where}
$$
U^\tau(t) \equiv \left(
\begin{array}{cc} U_+(t) & 0 \\ 0 & U_-^{-1}(t) \end{array}  \right)
$$
and $U_{\pm}(t)$ are evolution operators for the Hamiltonians $H_{\pm}(t)$:
$$
U_{\pm}(t) = \Texp (-i \int_0^T H_{\pm}(t) dt)
$$

Now we prove that
the eigenfunctions with nonzero eigenvalues
does not contribute in ${\ind}_B H$
and calculate Berry phases
of zero-modes.
The following theorem formalizes the statement:

\begin{theorema} \cite{Ch1}
\label{CASE}
\ Let's $H(t)$ is CASE-Hamiltonian on $[0,T]$ and \\
${\protect{\ind}}_W H(0) \ne 0$
then
\begin{equation}
{\ind}_B H \equiv  {\sDet}_{reg} U(T) =
\exp
\left(
-\int_0^T \sTr (B(t)P_0(t)) dt
\right)
\label{ind}
\end{equation}
where
$$
B(t) \equiv
\left(
       \begin{array}{cc}
              B_+(t) & 0
	      \\
	      0 & B_-(t)
	\end{array}
\right)
$$
and $P_0(t)$ is the projector on $\ker H(t)$.
\end{theorema}

{\it Proof.}
At first we prove that RHS is well-defined and independent on the choice of
instant bases in $\ker H_\pm(t)$ which keep their dimensions due to
Lemma \ref{invariant}.

\begin{lemma}
\label{trace}
$\ \exp(-\int_0^T \sTr (B(t)P_0(t)) dt)$
does not depend on the choice of
instant orthonormalized bases $\{\varphi^{\alpha}_{0\pm}(t)\}$ with cyclic
condition (\ref{cyc2}) in $\ker H_\pm(t)$.
\end{lemma}

{\it Proof.}
If we introduce new bases $\{\chi^{\alpha}_{0\pm}(t)\}$ in $\ker H_\pm(t)$
by the relations:
\begin{equation}
\chi_{0\pm}^{\alpha}(t) = \sum\limits_{\alpha^\prime=1}^{\dim\ker\bb{H}_\pm}
v_{\pm}^{\alpha\alpha^\prime}(t) \varphi_{0\pm}^{\alpha^\prime} (t)
\end{equation}
then $\Tr B(t)|_{\ker H_\pm(t)}$ can be calculated in new bases:
\begin{equation}
\begin{array}{l}
\Tr (B(t)|_{\ker H_\pm(t)})_\chi =
\sum\limits_{\alpha =1}^{\dim\ker\bb{H}_\pm}
\langle\chi_{0\pm}^{\alpha}(t) |\dot\chi_{0\pm}^{\alpha}(t)\rangle  =  \\
= \sum\limits_{\alpha,\beta,\gamma =1}^{\dim\ker\bb{H}_\pm}
\langle v_{\pm}^{\alpha\beta}(t)\varphi_{0\pm}^{\beta}(t) |
v_{\pm}^{\alpha\gamma}(t)\dot\varphi_{0\pm}^{\gamma}(t) +
\dot v_{\pm}^{\alpha\gamma}(t)\varphi_{0\pm}^{\gamma}(t) \rangle  =  \\
= \sum\limits_{\alpha,\beta,\gamma =1}^{\dim\ker\bb{H}_\pm} (
\langle\varphi_{0\pm}^{\beta}(t) | \dot\varphi_{0\pm}^{\gamma}(t)\rangle
\delta^{\beta\gamma} +
\delta^{\beta\gamma}{v_{\pm}^{-1}}^{\beta\alpha} \dot v_{\pm}^{\alpha\gamma}
) =  \\
= \Tr (B(t)|_{\ker H_\pm(t)})_\varphi +
\Frac{\partial}{\partial t} \ln\Det\| v_{\pm}^{\alpha\beta}(t)\|
\end{array}
\end{equation}
where we use the formula
\begin{equation}
\Frac{\partial}{\partial t} \Det V(t) =
\Det V(t) \Tr \Bigl(\Frac{\partial V(t)}{\partial t} V^{-1}(t)\Bigr)
\label{deriv}
\end{equation}
Hence
\begin{equation}
\sTr (B(t)|_{\ker H(t)})_\chi =
\sTr (B(t)|_{\ker H(t)})_\varphi +
\Frac{\partial}{\partial t} \ln\Frac{\Det\| v_{+}^{\alpha\beta}(t)\| }
{\Det\| v_{-}^{\alpha\beta}(t)\| }
\label{change}
\end{equation}
$\| v_{\pm}^{\alpha\beta}(T)\| $ is equal to
$\| v_{\pm}^{\alpha\beta}(0)\| $
due to the bases $\{\varphi^{\alpha}_{0\pm}(t)\}$ and
$\{\chi^{\alpha}_{0\pm}(t)\}$ obey the cyclic condition (\ref{cyc2}).
Therefore last term after the integration by $t$ from $0$ to $T$
gives $2\pi i k ,\ k\in Z$ due to the logarithm of complex number is
multivaluable function.
\eop

Now let's return to the proof of the Theorem \ref{CASE}. Keeping in mind
the regularization we can calculate the  time derivative
$\Frac{\partial}{\partial t} \Det U^{\tau}(t)$:
\begin{equation}
\begin{array}{c}
\Frac{\partial}{\partial t} \Det U^{\tau}(t) =
\Frac{\partial}{\partial t} (\Det U_+(t)\Det U_{-}^{-1}(t))      =  \\
= \Frac{\partial}{\partial t} \Det U_+(t) (\Det U_-(t))^{-1} +
\Det U_+(t)\Frac{\partial}{\partial t} (\Det U_-(t))^{-1}
\end{array}
\label{pr1}
\end{equation}
using formulae (\ref{deriv}) and (\ref{ev}) for the systems governed by
Hamiltonians $H_\pm(t)$ we get
\begin{equation}
\begin{array}{c}
\Frac{\partial}{\partial t} \Det U^{\tau}(t) =
\\
= \Det U^{\tau}(t)
\Bigl(
       -\Tr (B_+(t)+iE_+(t)) +\Tr (B_-(t)+iE_-(t))
\Bigr)
=
\\
= - \Det U^{\tau}(t) \sTr B(t)
\end{array}
\label{pr2}
\end{equation}
We can solve this equation.
Taking into account the initial condition $\Det U^{\tau}(0)=1$:
\begin{equation}
\Det U^{\tau}(T)\left. \over \right|_{E([0,\lambda])(t)\bb{H}} =
\exp
\Bigl(
-\int^T_0 \sTr B(t)
\left. \over \right|_{E([0,\lambda])(t)\bb{H}}dt
\Bigr)
\label{pr3}
\end{equation}
Due to CASE-condition (\ref{SSev}) RHS of this equation has
$\lambda\to\infty$ limit. Therefore
\begin{equation}
{\sDet}_{reg} U(T) =
\Det U^{\tau}(T) \left. \over \right|_{P_{0}(T)\bb{H}}  =
\exp(-\int_0^T \sTr B(t) \left. \over \right|_{P_0(t) \bb{H}} dt)
\end{equation}
Thus we have proved Theorem \ref{CASE} \eop

{\bf Note 3.1}
It is possible to generalize Theorem \ref{CASE} on the case of
CASE-Hamiltonians with ${\ind}_W H(t)=0$.
Then in the statement we
have to replace $\exp(-\int_0^T (B(t)P_0(t))dt)$ by
$\exp(-\int_0^T (B(t)\tilde P_0(t))dt)$ where $\tilde P_0(t)$ is projector
on eigenspace with eigenvalue which somewhere on interval $[0,T]$ comes in
zero.

{\bf Note 3.2}
For the case of 1-dimensional
$\ker\bb{H}_{\pm}$ index $\ind_B H$ gives us no more
than $\exp(\Delta\varphi)$ where $\Delta\varphi$ is a difference of Berry
phases of zero-modes in "bosonic" and "fermionic" spaces.
In this form
${\ind}_B$ is analog of supersymmetric scattering index \cite{BMS} which
also calculates the difference of phases (scattering phases) in
"bosonic" and "fermionic" spaces.

In general case the Supersymmetric Berry Index is a complex number
on the unit circle and this number can be changed via the variation
of the CASE-Hamiltonian.
It would be especially interesting to describe
some situations for which this number has to be in the
set of discrete numbers.
Now we formulate the simplest condition on the CASE-Hamiltonian
which leads to the discretness of possible Supersymmetric Berry Indeces.

Let's Hilbert space $\bb{H}$ has additional structure which we will
call "conjugation".
Formally it means that there is the involution
$P : \bb{H} \rightarrow \bb{H}$ such that for
$\forall\varphi,\psi\in\bb{H}$ and $c\in C$
\begin{enumerate}
\item $ P^{2} = I $
\item $ P c\varphi = \overline{c} P \varphi $
\item
$ \langle P \varphi | P \psi \rangle =
\overline{\langle \varphi|\psi \rangle} $
\end{enumerate}

\begin{theorema}
\label{real}
\cite{Ch1}
If the "conjugation" $P$ is consistent with the supersymmetric
involution i.e. $[P,\tau] = 0$
and CASE-Hamiltonian obeys the condition:
$$
H(t) P = P H(t)   \qquad \forall t \in [0,T]
$$
then
$$ {\ind }_B H(t) = \pm 1 $$

{\bf Note 4.1} In this form the Theorem \ref{real} generalize well-known
fact \cite{Simon}
that Berry phase of real Hamiltonian is equal zero.
\end{theorema}

{\it Proof.}
{}From the facts that the "conjugation" $P$  commutes with the Hamiltonian
and $P\bb{H}_\pm = \bb{H}_\pm$
immediately follows that for $\forall t \in [0,T]$ it is possible to choose
"real" instant eigenfunctions of the operators $H_\pm(t)$ i.e.
such that $P|\chi^\alpha (t)\rangle = |\chi^\alpha  (t)\rangle$.
We also suppose the orthonormalization condition for the function.
It leads to the next property of the instant eigenfunctions:
$$
\langle\chi^\alpha (t)|\chi^\beta (t^{\prime})\rangle =
\langle P \chi^\alpha (t)|P \chi^\beta (t^{\prime})\rangle =
\overline{\langle\chi^\alpha (t)|\chi^\beta (t^{\prime})\rangle} \in R
$$
and therefore
$\langle\chi^\alpha (t)|\dot\chi^\beta (t^{\prime})\rangle \in R$
for all moments $t, t^{\prime } \in [0,T]$.

However in general case $\chi_\alpha (T) \ne \chi_\alpha (0)$
Using formulae (\ref{ind},\ref{change}) we get following expression
for ${\ind}_B H$:
$$
\begin{array}{c}
{\ind}_B H =
\exp\left(
          \int\limits_0^T
          \sum\limits_{\pm}\pm
	  \sum\limits_{\alpha =1}^{\dim\ker\bb{H}_\pm}
              \langle\chi_{0\pm}^{\alpha}(t) |
                      \dot\chi_{0\pm}^{\alpha}(t)
	      \rangle
	   dt +
           \ln\left.
	          {\Det\| v_{+}^{\alpha\beta}(t)\| }
                  \over
	          {\Det\| v_{-}^{\alpha\beta}(t)\| }
	       \right|_0^T
    \right)
\end{array}
$$
where $v_{\pm}^{\alpha\beta}(t)$ is the transfer matrix from
basis $\{\chi_{0\pm}^{\alpha}(t)\}$ to basis
$\{\varphi_{0\pm}^{\beta}(t)\}$ in $\ker H_{\pm}(t)$.
$\langle\chi_{0\pm}^{\alpha}(t) |\dot\chi_{0\pm}^{\alpha}(t)\rangle =0 $
because on the one hand it is real but on another hand it is imaginary
due to normalization condition on $\chi_{0\pm}^{\alpha}(t)$. So
\begin{equation}
{\ind}_B H =
\Frac{\Det(\| v_{+}^{\alpha\beta}(0)\| ^{-1} \| v_{+}^{\beta\gamma}(T)\| )}
{\Det(\| v_{-}^{\alpha\beta}(0)\| ^{-1}\| v_{-}^{\beta\gamma}(T)\| )}
\label{p1}
\end{equation}
Due to the cyclic condition (\ref{cyc2}) the matrices
$\| v_{\pm}(0)\| ^{-1} \| v_{\pm}(T)\| $ are the transfer matrices from
$\{\chi_{0\pm}^{\alpha}(T)\}$ to $\{\chi_{0\pm}^{\alpha}(0)\}$:
$$
\| v_{\pm}(0)\| ^{-1} \| v_{\pm}(T)\| ^{\alpha\beta} =
\langle\chi_{0\pm}^{\beta}(T) |\chi_{0\pm}^{\alpha}(0)\rangle \in R
$$
Matrices $\| v_{\pm}(0)\| ^{-1} \| v_{\pm}(T)\| $ are unitary and real.
Therefore both determinant in (\ref{p1}) are equal to $\pm 1$
and ${\ind}_B H = \pm 1$.
\eop

\section{Examples}
It is well-known how important to find the simple example which illustrates
the general structure and is not shaded the treatment by long calculations.
As such example we can consider the
supersymmetric harmonic oscillator on complex plane.
\subsection{Supersymmetric harmonic oscillator on complex plane.}

        The supersymmetric harmonic oscillator on a complex plane is the
        simplest example of the suggested in \cite{J}  Holomorphic
        Supersymmetric Quantum Mechanics which was investigated in many
	papers~\cite{HSQM,Klimek,AO}.
	It was shown that the Hamiltonian of supersymmetric
	harmonic oscillator has pure point spectrum and there is the
	only single zero mode in "fermionic" subspace. So Witten index for
	such operator is equal to $-1$.  Now we demonstrate that this
	Hamiltonian can be regarded as CASE-one and calculate the
	Supersymmetric Berry Index.  For this supersymmetric system the
	supercharge $Q(t)$ has the form:
$$
Q(t) = \sqrt{2}\left( \begin{array}{cccc}
0 & 0 & \bar c(t)\bar z & -\Frac{\partial}{\partial z}  \\
0 & 0 & \Frac{\partial}{\partial \bar z} & - c(t) z \\
c(t) z & -\Frac{\partial}{\partial z} & 0 & 0 \\
\Frac{\partial}{\partial \bar z} & -\bar c(t)\bar z & 0 & 0 \\
\end{array} \right) \ ,
$$
        The corresponding supersymmetric Hamiltonian
        $H(t) = Q^2 (t)$ can be put down:
$$
H(t) = \left( \array{cc}
		H_+(t) &  0
		\\
		0   &  H_-(t)
	      \endarray
	\right) \ ,
$$
where
$$
H_+(t) = 2 \left( \begin{array}{cc}
|c(t)z|^2 - \Frac{\partial^2}{\partial z\partial\bar z} & 0 \\
0 & |c(t)z|^2 - \Frac{\partial^2}{\partial z\partial\bar z}
\end{array} \right) \ ,
$$
$$
H_-(t) = 2 \left( \begin{array}{cc}
|c(t)z|^2 - \Frac{\partial^2}{\partial z\partial\bar z} & c(t) \\
\bar c(t) & |c(t)z|^2 - \Frac{\partial^2}{\partial z\partial\bar z}
\end{array} \right) \ ,
$$
        and together with supersymmetric involution
$
\tau =
\left(
\array{cc}
I &  0
\\
0 & -I
\endarray
\right)
$
        forms the SQM-algebra.
        In the presented formulae time dependence appears through the
        arbitrary smooth function $c(t)$ such that $c(0) = c(T)$ and
        $c(t) \ne 0 \ \mbox{for}\ \forall t \in [0,T]$.
\begin{lemma}
\label{oscil}
$H(t)$ is CASE-Hamiltonian
\end{lemma}

{\it Proof.}
Indeed, it is sufficient to write down
$ [\dot Q(t),{Q}(t)]\left. \over \right|_{\bb{H}_+} $:
\begin{equation}
\begin{array}{l}
\left. [\dot Q(t), Q(t)] \right|_{\bb{H}_+} = \\
=2 \left( \begin{array}{cc}
\overline{\dot c(t)}c(t)|z|^2 - \overline{c(t)}\dot c(t)|z|^2 &
-2 \overline{\dot c(t)z} \Frac{\partial}{\partial z} \\
-2 {\dot c(t)z} \Frac{\partial}{\partial\bar z} &
\overline{c(t)}\dot c(t)|z|^2 - \overline{\dot c(t)}c(t)|z|^2
\end{array} \right)
\end{array}
\label{3}
\end{equation}
and to take eigenfunctions of $H_+(t)$ in the form
$$
\varphi_{j+}^{2\alpha-1} =
\left(
   \array{c}
   \xi_j(t) \\ 0
   \endarray
\right) ,
\qquad
\varphi_{j+}^{2\alpha} =
\left(
   \array{c}
   0 \\ \xi_j(t)
\endarray
\right)   .
$$
The diagonal terms of the matrix
$\left. [\dot Q(t),Q(t)]\right|_{\bb{H}_+} $
has the opposite signs. Therefore
$$
\Sum_\alpha
          \langle
          \varphi_{j+}^\alpha
          |  [\dot Q(t),Q(t)]  |
          \varphi_{j+}^\alpha
          \rangle
= 0
\  \forall j \ \mbox{such that}\  E_j(t)>0
$$
This leads to the required equality (\ref{SSev}) for the trace.
\eop

{\bf Proposition 1}

$$
\ind_B H(t) = (-1)^{\ind\limits_{[0,T]} c(t)}
$$

{\it Proof.}
Due to the fact that $H(t)$ is CASE-Hamiltonian with $\ind_W H(t)$
$ = -1$
and single zero mode is in "fermionic" sector \cite{J}
we can apply the Theorem \ref{CASE} for calculation of $\ind_B H(t)$.

        Using the notation $c(t) = r(t)\exp(i\theta(t)) $ and the explicit
        form of the zero-mode
$$
\varphi_{0-}(t) =
\sqrt{\Frac{r(t)}{\pi}}\exp(-r(t)|z|^2)
\left(
\array{c}
\exp (i\theta(t)) \\ -1
\endarray
\right)
$$
we compute $\sTr (B(t)P_0)$:
\begin{eqnarray}
- \sTr B(t) \left. \over \right|_{\ker H} =
\langle \varphi_{0-}(t)|\dot\varphi_{0-}(t)\rangle  =
\nonumber\\
={\Frac{r(t)}{\pi}}\int_C \exp(-2r(t)|z|^2)
(i\dot\theta(t) + \Frac{\dot r(t)}{r(t)} - 2\dot r(t)|z|^2) d\bar zdz =
\Frac{i}{2}\dot\theta(t)
\end{eqnarray}
        According to the Theorem \ref{CASE} Supersymmetric Berry Index is
        equal to
$$
{\ind}_B H =
\exp\left( \Frac{i}{2}\int\limits_{0}^{T}\dot\theta (t) dt \right)
= (-1)^{\ind\limits_{[0,T]} c(t)} \qquad \eop
$$

        This complete the consideration of Supersymmetric
        Harmonic Oscillator on complex plane.

Now we consider the general case of the Supersymmetric Meromorphic Quantum
Mechanics on the Riemann surface. This system contains the previous example
as a particular case.

\subsection{Meromorphic Supersymmetric Quantum
Mechanics on Riemann surface}

Let's define SQM on the arbitrary genus compact Riemann surface
$ M_{0} $ with meromorphic superpotential with poles in
$ z_{1},\ldots ,z_{n} \in M_{0}$.
To do this K\"ahler metric $ g $ which is euclidean at infinity
in the points $ z_{1},\ldots ,z_{n}$ was introduced \cite{SQMR}.

For this metric there are open neigh\-bor\-hoods
$O_{R_{i}}$ of $z_{i}$ and diffeomorphic maps $\phi _{i} $ of $O_{R_{i}}
\setminus \{z_{i} \} $ to open sets $ CB_{R_{i}}=\{ u\in C:| u |
>R_{i}\} $ on complex plane such that on each $ O_{R_{i}}$ the metric is
the pullback by $\phi _{i}$ of the euclidean metric on $CB_{R_{i}}$.

Hilbert space is that of square integrable differential forms:
$\bb{H} \equiv \Lambda_{2}(M)$ with scalar product
\be
\langle \omega|\phi \rangle  = \int_{M}\overline\omega \wedge {*\phi}
\label{scal}
\vspace{0.6cm}
\ee
where $*$ is Hodge operator.

We will consider the time-depended
meromorphic superpotential $F(z,t)$ such that its poles
$ z_{1},\ldots ,z_{n} \in M_{0}$
are independent on $t$ and $F(z,T)=F(z,0)$.

Supercharges, the supersymmetric involution and the
Hamiltonian were defined as the closure in $\bb{H}$
of correspondent operators defined on $C^{\infty}_{0}(M)$-
forms by formulae:
\be
\begin{array}{l}
Q_{+}(t) =
\Frac{\partial}{\partial
\bar {z}}d\bar
z\wedge
+ F_{z}(z,t) dz\wedge \ ,
\qquad
Q_{-}(t) = \bigl(Q_{+}(t)\bigr)^{*} \ ,
\\
{ }
\\
\tau = (-1)^{\mathop{\mbox{\rm N}}} \ ,
\qquad
Q(t) = Q_+(t) + Q_-(t) \ ,
\qquad
H(t) = (Q(t))^2
\end{array}
\ee
where $\mathop{\mbox{\rm N}}$ is the degree of the form.
The index $z$ stands for derivative on $z$.
These operators obey SQM relations (\ref{SQM}).
In the work \cite{SQMR} it was shown that
operator $H(t)$ has compact resolvent and hence pure discrete spectrum.

Let's choose the basis of differential forms:
$1, gd\bar{z}\wedge dz/2, dz, d\bar z$.
The first two forms belong to $\bb{H}_+$,
the two latter forms belong to $\bb{H}_-$.
In this basis all operators can be represented in matrix form
(for the sake of simplicity we omit arguments $z,t$ of function $F$):
\begin{equation}
Q(t) =
\left(
        \begin{array}{cccc}
              0 & 0 & \Frac{2}{g}\overline{F_z} &
              -\Frac{2}{g}\Frac{\partial}{\partial z}
	\\
	      0 & 0 & \Frac{2}{g}\Frac{\partial}{\partial\bar z} &
	      -\Frac{2}{g}F_z
	\\
	      F_z & -\Frac{\partial}{\partial z} & 0 & 0
	\\
              \Frac{\partial}{\partial\bar z} & -\overline{F_z} & 0 & 0
	\\
        \end{array}
\right) \ ,
\end{equation}
\begin{equation}
\begin{array}{c}
	H_+(t) =
	      \left(
		\begin{array}{cc}
      		       \Frac{2}{g}
      		       (|F_z|^2 -
		       \Frac{\partial^2}{\partial z\partial\bar z})
		       & 0
	         \\
		       0 & \Frac{2}{g}
		       (|F_z|^2 -
		       \Frac{\partial^2}{\partial z\partial\bar z})
		\end{array} \right) \ ,
\\ { } \\
H_-(t) =
\left(
	    \begin{array}{cc}
	    	-\Frac{\partial}{\partial z} \Frac{2}{g}
	    	\Frac{\partial}{\partial\bar z} +
         	\Frac{2}{g}|F_z|^2 &
     	        \Frac{2}{g}F_{zz}
	    \\
	    	\Frac{2}{g}
	    	\overline{F_{zz}} &
	    	-\Frac{\partial}{\partial\bar z}
	    	\Frac{2}{g}\Frac{\partial}{\partial z} +
	    	\Frac{2}{g}
	    	|F_z|^2
           \end{array}
\right)
\ .
\end{array}
\end{equation}
and
\begin{equation}
[\dot Q(t), Q(t)]
\left. \over \right|_{\bb{H}_+} =
\Frac{2}{g}
\left(
        \begin{array}{cc}
		\dot{\overline{F}}_z F_z
		- \overline{F}_z \dot F_z &
		-2 \dot{\overline{F}}_z
		\Frac{\partial}{\partial z}
	\\
		-2 {\dot F_z}
		\Frac{\partial}{\partial\bar z} &
		\overline{F}_z\dot F_z
		-\dot{\overline{F}}_z F_z
	\end{array}
\right)  \ .
\end{equation}

In the subspace $\bb{H}_+$ we can take eigenfunction in the form of
$$
\varphi^{2\alpha-1}_{j+}(t) =
\left(
\begin{array}{c} \xi^{\alpha}_j(t)
\\
0
\end{array}
\right)  \ ,
\qquad
\varphi^{2\alpha}_{j+}(t) =
\left(
\begin{array}{c}
0
\\
\xi^{\alpha}_j(t)
\end{array}
\right)  \ ,
$$
and diagonal terms of matrix
$[\dot Q(t), Q(t)]|_{\bb{H}_+}$ are opposite in
the sign therefore
the condition (\ref{SSev}) holds true in this case
due to the same
reason as in the previous subsection .

If $F(z,t)$ depends on $t$ such that $H(t)$ obeys adiabatic
condition (\ref{Ad})
then the operator $H(t)$ is CASE-Hamiltonian.
Therefore ${\ind}_B H$ exists.
One can show that the Hamiltonian in question satisfies the conditions of
the Theorem \ref{real}
(for usual complex conjugation as the involution $P$)
and hence ${\ind}_B H$ is equal to $\pm 1$.

To escape the technical difficulties we calculate
the index for particular case of $F(z,t)=\exp(i\theta(t))f(z)$,
$\theta(T) = \theta(0) + 2\pi L,\ L\in Z$.

{\bf Proposition 2} \\
{\it Let
$\chi (M_{0})$ is Euler cha\-rac\-te\-ris\-tics of compact Riemann
surface $M_{0}$ and $D$ is a divisor of poles of the differential
$F_zdz$,
then
$$
{\ind}_B H = (-1)^{L(\chi (M_{0}) + \deg D)}
$$
}
{\it Proof. }
According to the Theorem \ref{CASE} we have to
investigate zero-modes.  In the work \cite{SQMR} number of zero-modes of
Hamiltonians $H_{\pm}(t)$ has been calculated and it was shown that the
Hamiltonian $H_+(t)$ has no zero-modes, $H_-(t)$
has $K\equiv \chi (M_{0}) + \deg D$ zero-modes.

Given the basis of subspace of zero-mode at moment $t=0$
$$
\varphi_{0-}^\alpha(0) =
\left(
 \begin{array}{c}
    \xi_1^\alpha  \\  \xi_2^\alpha
 \end{array}
\right)
$$
we can construct instant bases for any $t\in [0,T]$:
$$
\varphi_{0-}^\alpha(t) =
\left(
\begin{array}{c}
     \exp(i\theta(t))\xi_1^\alpha
     \\
     \xi_2^\alpha
\end{array}
\right)
$$
\begin{equation}
{\ind}_B H =
\exp
\left(
      \int_0^T \sum\limits_{\alpha =1}^K
\langle
      \varphi_{0-}^\alpha(t)
|
      \dot\varphi_{0-}^\alpha(t)
\rangle
      dt
\right) =
\int_0^T
i\dot\theta(t) dt
\sum\limits_{\alpha =1}^K
\| \xi_1^\alpha\| ^2
\end{equation}
At the moment $t=0$ we can take real eigenbasis. Then
$\| \xi_1^\alpha\| ^2 = \| \xi_2^\alpha\| ^2 = 1/2$.
Therefore
$
{\ind}_B H = (-1)^{L(\chi (M_{0}) + \deg D)}
$
\eop

The Proposition 2 generalizes the Proposition 1 of the previous section
on the case of the Meromorphic Supersymmetric Quantum Mechanics.

\section{Conclusion remarks}

In this paper we investigated the possibility to insert the concept
of topological phases of cyclic adiabatic evolution (Berry phases) to
the framework of Supersymmetric Quantum Mechanics and introduced on
this basis new topological index -- Supersymmetric Berry Index.
To illustrate the scheme this index
was calculated for the Holomorphic and Meromorphic Supersymmetric
Quantum Mechanics. For these cases index is connected with winding number
of parametric function which generates the adiabatic evolution.

However at the end of the paper we would like to outline some
directions of developments.

We discussed in the paper only one way to calculation the index -
through the explicit computing of the contributions of zero-modes.
The second step is to generalize the Index Theorems the sense
discussed in the Introduction,
namely is to represent the CASE-index in
functional
integral form that allow us to treat
Supersymmetric Berry phase  using quantum field theory methods.

The Generalized Supersymmetry has proved that it is additional powerful
tool of mathematical investigation into the framework of supersymmetry.
So to our mind it is interesting to generalize the treatment of
Supersymmetric Berry Phase to case of Generalized Supersymmetry.

The last point is the looking for other possibilities of the discretness
of the index because this is a straightforward way to the real topological
stability. For instance it is interesting to find the conditions
under which
the index takes the value in the set of complex roots of unit.

We are going to return to these questions in the next papers.

\begin{center}
{\bf Acknowledgment}
\end{center}
Additional partial support of the work (K.N.I) was received through Euler
Stipend of German Mathematical Society, EC Grant INTAS N 939
and Grant of Russian Fund of the Fundamental
Investigations N $94-01-01157-a$.

\end{document}